\documentclass[aip,jcp,reprint,superscriptaddress,longbibliography]{revtex4-2}

\usepackage{amsmath}
\usepackage{amssymb}
\usepackage{array}
\usepackage{bm}
\usepackage{graphicx}
\usepackage{dcolumn}
\usepackage{hyperref}
\hypersetup{hidelinks}
\usepackage{xcolor}
\usepackage{tikz}
\usetikzlibrary{arrows.meta,positioning}

\newcommand{\Skala}{\textsc{Skala}}
\newcommand{\GauXC}{\textsc{GauXC}}
\newcommand{\CPtwoK}{\textsc{CP2K}}
\newcommand{\PySCF}{\textsc{PySCF}}
\newcommand{\GPW}{\textsc{GPW}}
\newcommand{\GAPW}{\textsc{GAPW}}
\newcommand{\GTH}{\textsc{GTH}}
\newcommand{\GPWGTH}{\GPW{}-\GTH{}}
\newcommand{\GAPWAE}{\GAPW{}-AE}
\newcommand{\GAPWGTH}{\GAPW{}-\GTH{}}
\newcommand{\GAPWECP}{\GAPW{}-ECP}
\newcommand{\SITableFont}{\normalsize}
\newcommand{\Tr}{\operatorname{Tr}}
\newcommand{\dd}{\mathrm{d}}

\begin{document}

\title{Molecular Implementation of the Machine-Learned \Skala{} Exchange--Correlation Functional in \CPtwoK{} through \GauXC{}}

\author{Franz P\"oschel}
\thanks{These authors contributed equally to this work.}
\affiliation{Center for Advanced Systems Understanding (CASUS), G\"orlitz, Germany}
\affiliation{Helmholtz-Zentrum Dresden-Rossendorf (HZDR), Dresden, Germany}

\author{Johann Pototschnig}
\thanks{These authors contributed equally to this work.}
\affiliation{Center for Advanced Systems Understanding (CASUS), G\"orlitz, Germany}
\affiliation{Helmholtz-Zentrum Dresden-Rossendorf (HZDR), Dresden, Germany}

\author{Frederick Stein}
\affiliation{Center for Advanced Systems Understanding (CASUS), G\"orlitz, Germany}
\affiliation{Helmholtz-Zentrum Dresden-Rossendorf (HZDR), Dresden, Germany}

\author{Andreas Knüpfer}
\affiliation{Center for Advanced Systems Understanding (CASUS), G\"orlitz, Germany}
\affiliation{Helmholtz-Zentrum Dresden-Rossendorf (HZDR), Dresden, Germany}

\author{Thijs Vogels}
\affiliation{Microsoft Research AI for Science, Amsterdam, Netherlands}

\author{Stefano Battaglia}
\affiliation{Microsoft Research AI for Science, Amsterdam, Netherlands}

\author{Sebastian Ehlert}
\affiliation{Microsoft Research AI for Science, Berlin, Germany}

\author{J\"urg Hutter}
\affiliation{Department of Chemistry, University of Zurich, Zurich, Switzerland}

\author{Thomas D. K\"uhne}
\email{tkuehne@cp2k.org}
\affiliation{Center for Advanced Systems Understanding (CASUS), G\"orlitz, Germany}
\affiliation{Helmholtz-Zentrum Dresden-Rossendorf (HZDR), Dresden, Germany}
\affiliation{Institute of Artificial Intelligence, Technische Universit\"at Dresden, Dresden, Germany}

\date{\today}

\begin{abstract} Machine-learned exchange--correlation (XC) functionals offer a route to improve Kohn--Sham density-functional theory without incurring the cost of explicitly correlated electronic-structure methods. Their use in production simulation codes, however, requires a well-defined mapping between the learned model and the host-code density representation. We formulate and implement a \Skala{}-1.1 interface in \CPtwoK{} through the external \GauXC{} library. \CPtwoK{} supplies the geometry, Gaussian basis, spin-resolved atomic-orbital density matrix, and communicator, while \GauXC{} evaluates the XC energy, atomic-orbital potential matrix, and available nuclear derivatives. The interface accepts both all-electron and valence-only density matrices. The latter may arise from separable dual-space pseudopotentials or molecular effective-core potentials. Implementation errors are isolated from functional differences by comparing the Perdew--Burke--Ernzerhof (PBE) functional evaluated through \GauXC{} with native \CPtwoK{} PBE. The resulting interface gives consistent energies, forces validated against finite-difference total-energy checks, and force-based molecular-virial diagnostics for representative molecular cases. The dietGMTKN55 benchmark suite is evaluated with an all-electron Gaussian augmented plane-wave treatment for elements up to bromine and def2 effective-core potentials for the heavier elements. The resulting aggregate mean absolute deviation of \(1.255~\mathrm{kcal\,mol^{-1}}\) is within \(0.012~\mathrm{kcal\,mol^{-1}}\) of the corresponding \Skala{} reference value of \(1.243~\mathrm{kcal\,mol^{-1}}\). This work establishes a validated molecular implementation of \Skala{} in \CPtwoK{} through \GauXC{}. \end{abstract}

\maketitle

\section{Introduction}

Kohn--Sham (KS) density-functional theory (DFT) is the default electronic-structure framework for many simulations of molecules, liquids, and materials because it offers a favorable balance between accuracy and computational cost.\cite{Hohenberg1964,Kohn1965}
In practical simulations, however, the exchange--correlation (XC) functional remains the principal uncontrolled approximation.
Semilocal functionals such as the Perdew--Burke--Ernzerhof (PBE) generalized-gradient approximation (GGA)\cite{Perdew1996} are robust, inexpensive, and naturally compatible with pseudopotentials and analytical forces.
The same locality that makes them so useful also limits their representational flexibility: the XC energy density is constrained to depend on local ingredients such as the density, its gradient, and sometimes the kinetic-energy density.

Systematically more accurate post-Hartree--Fock (post-HF) approaches are available in \CPtwoK{} for increasingly realistic condensed-phase settings.
Examples include resolution-of-the-identity second-order M{\o}ller--Plesset perturbation theory (RI-MP2) and spin-unrestricted MP2 forces,\cite{Rybkin2016} double-hybrid density functionals with gradients, stress tensors, and auxiliary density matrix method (ADMM) acceleration,\cite{Stein2022} low-scaling sparse-tensor Hartree--Fock and correlated-gradient machinery with graphics processing unit (GPU) acceleration,\cite{Bussy2023} and massively parallel random phase approximation (RPA) gradients for molecular crystals.\cite{Stein2024}
These developments complement the cubic-scaling RPA and Green's-function-based \(GW\) infrastructure in the Gaussian and plane-wave (\GPW{}) framework\cite{Wilhelm2016b,Wilhelm2018} and related high-accuracy correlation approaches, including the original \(\sigma\)-functional and its recent \CPtwoK{} implementation.\cite{Trushin2021Sigma,Mandalia2025SigmaCP2K}
Yet even with sparsity, ADMM, optimized tensor contractions, GPU acceleration, and Message Passing Interface (MPI)/Open Multi-Processing (OpenMP) parallelization, these methods remain too expensive for routine large-cell ab initio molecular dynamics (AIMD),\cite{Kuehne2007PRL,Kuehne2014WIREs} high-throughput DFT-verification workflows,\cite{Bosoni2024NatRevPhys} or the large force-and-energy data sets needed to train transferable machine-learning models.
Machine-learned XC functionals address this gap by aiming for accuracy beyond that of traditional semilocal functionals while retaining a computational cost suitable for AIMD production sampling and high-throughput DFT workflows.

\Skala{} is a recently introduced deep-learning-based XC functional that attains accuracy competitive with state-of-the-art hybrid functionals across main-group chemistry at a cost closer to semilocal DFT.\cite{Skala2025}
Its public software distribution provides an implementation through \GauXC{}, allowing the same differentiable model to be used by multiple electronic-structure codes.\cite{Williams2020}
This portability requires an unambiguous host--library interface that defines the density representation, spin convention, molecular quadrature, XC energy and potential, and nuclear derivatives.

\CPtwoK{} is a natural target for such an interface because its \textsc{Quickstep} module represents the Kohn--Sham orbitals in a localized Gaussian basis and uses auxiliary plane-wave (PW) grids for efficient electrostatic and native XC operations.\cite{Lippert1997,CP2K2020,CP2KMadeSimple2026}
For the integration of \Skala{} in \CPtwoK{}, three classes of molecular calculations and their corresponding AO densities are relevant.
The \GPW{} method with norm-conserving Goedecker--Teter--Hutter (\GTH{}) separable dual-space pseudopotentials treats the valence electrons explicitly and represents the core effects through the pseudopotential Hamiltonian.
The Gaussian augmented plane-wave (\GAPW{}) method with \texttt{POTENTIAL ALL}, denoted \GAPWAE{} below, uses an all-electron Hamiltonian and an atomic-orbital (AO) density matrix that represents the complete electron density.
Pseudopotential \GAPW{} calculations instead use either a \GTH{} potential or an effective-core potential (ECP), so their AO density matrix represents only the electrons retained explicitly by the corresponding valence Hamiltonian.  ECPs are particularly relevant for heavier elements in standard molecular basis-set families such as def2.\cite{Weigend2005,Andrae1990}

The same interface is used for all three classes: \CPtwoK{} passes a spin-resolved AO density matrix to \GauXC{}, and \GauXC{} returns the XC energy, the AO XC potential matrix, and available nuclear derivatives.  We validate this interface first with PBE, comparing the results obtained through \GauXC{} against the native \CPtwoK{} PBE implementation and then with \Skala{} finite-difference checks and dietGMTKN55, a representative subset of the General Main Group Thermochemistry, Kinetics, and Noncovalent Interactions (GMTKN55) database.\cite{Goerigk2017,Gould2018,Skala2025}

\section{Density Representations in \CPtwoK{}}

\begingroup

For a spin channel \(\sigma\), the molecular AO density matrix and the corresponding real-space density are
\begin{align}
  P_{\mu\nu}^{\sigma}
  &=
  \sum_i f_{i\sigma} C_{\mu i}^\sigma C_{\nu i}^\sigma,
  \\
  \rho_{\sigma}(\mathbf r)
  &=
  \sum_{\mu,\nu=1}^{N_\text{AO}}
  P_{\mu\nu}^{\sigma}
  \chi_{\mu}(\mathbf r)\chi_{\nu}(\mathbf r).
  \label{eq:ao_density}
\end{align}
Here, \(f_{i\sigma}\) is the occupation of molecular orbital \(i\).  The AO density matrix is the primary representation supplied to the molecular \GauXC{} interface.  For native \GPW{} and \GAPW{} operations, \CPtwoK{} additionally represents the valence density or the smooth component of the augmented density in an auxiliary PW basis.
\begin{equation}
  \widetilde{\rho}_{\sigma}(\mathbf r)
  =
  \frac{1}{\Omega}
  \sum_{\mathbf G}
  \widetilde{\rho}_{\sigma}(\mathbf G)
  e^{i\mathbf G\cdot\mathbf r},
  \label{eq:pw_density}
\end{equation}
This representation permits efficient evaluation of density-dependent energy terms, including \(E_{\mathrm H}[\rho]\) and \(E_{\mathrm{xc}}[\rho]\), and is particularly convenient for periodic boundary conditions.  In Eq.~\eqref{eq:pw_density}, \(\Omega\) is the simulation-cell volume.  The coefficients \(\widetilde{\rho}_{\sigma}(\mathbf G)\) are obtained from the AO density matrix by collocating products of Gaussian basis functions on uniform real-space grids and applying a fast Fourier transform.  In \GPW{} this auxiliary expansion represents the complete valence pseudo-density, whereas in \GAPW{} it represents the smooth component of an augmented density decomposition.\cite{Lippert1997,Lippert1999,CP2K2020}

The density supplied to \GauXC{} is reconstructed directly from the AO representation rather than from the auxiliary PW representation.  Its electronic content is determined by the Hamiltonian used in the calculation.
\begin{equation}
  \rho_{\sigma}^{\mathrm{in}}(\mathbf r)
  =
  \begin{cases}
    \rho_{v,\sigma}^{\mathrm{AO}}(\mathbf r),
      & \text{valence Hamiltonian},\\[2pt]
    \rho_{\mathrm{AE},\sigma}^{\mathrm{AO}}(\mathbf r),

      & \text{all electron}.
  \end{cases}
  \label{eq:input_density}
\end{equation}

The first case comprises \GPWGTH{}, \GAPWGTH{}, and \GAPWECP{}, whereas the second is realized by \GAPWAE{}.  Thus, the input density always contains exactly the electrons represented explicitly by the selected Hamiltonian.  Its integrated particle number follows directly from the AO overlap matrix.

\begin{equation}
  N_{\mathrm e}
  =
  \sum_{\sigma}\int
  \rho_{\sigma}^{\mathrm{in}}(\mathbf r)\,\dd\mathbf r
  =
  \sum_{\sigma}\Tr\!\left[\mathbf P^{\sigma}\mathbf S\right].
  \label{eq:particle_number}
\end{equation}

For pseudopotential and effective-core Hamiltonians, this trace gives the number of explicitly treated valence electrons.  For \GAPWAE{}, it gives the total number of electrons.  \GauXC{} receives the spin-resolved AO density matrix, molecular geometry, and basis-set information and evaluates the density, its derivatives, and the XC functional on an atom-centered molecular quadrature.  The auxiliary PW grid and the \GAPW{} one-center densities used internally by \CPtwoK{} are therefore not part of the molecular \GauXC{} interface.

\subsection{Gaussian and plane-wave approach}

The \GPW{} approach combines an atom-centered Gaussian representation of the Kohn--Sham orbitals with an auxiliary PW representation of the electronic density.\cite{Lippert1997,CP2K2020}  In practice, the density is collocated on a hierarchy of uniform real-space grids and transformed to reciprocal space, where the Hartree potential can be evaluated efficiently using fast Fourier transforms.  Semilocal XC contributions are likewise evaluated on the real-space grids.  \GTH{} pseudopotentials remove the core electrons from explicit treatment and yield smoothly varying valence pseudo-orbitals.\cite{Goedecker1996,Mirhosseini2026UZH}

For the \GauXC{} evaluation, the auxiliary PW mapping is bypassed and the same valence density is reconstructed directly from Eq.~\eqref{eq:ao_density} on the molecular quadrature.  Because the core electrons are absent from the Hamiltonian, this valence density is the complete density from which the XC energy and AO potential matrix are evaluated.

\subsection{Gaussian and augmented-plane-wave approach}

A direct PW representation of an all-electron density would require a prohibitively large cutoff because of its rapid variation close to the nuclei.  \GAPW{} avoids this difficulty by replacing the single auxiliary PW expansion with a smooth global density supplemented by localized one-center corrections.\cite{Lippert1999}  Omitting the spin label for clarity, the decomposition is
\begin{equation}
  \rho(\mathbf r)
  =
  \widetilde{\rho}(\mathbf r)
  +
  \sum_A
  \left[
    \rho_A^{\mathrm{hard}}(\mathbf r)
    -
    \rho_A^{\mathrm{soft}}(\mathbf r)
  \right],
  \label{eq:gapw_density}
\end{equation}

The smooth density is represented on the auxiliary grid, whereas the hard-minus-soft terms restore the rapidly varying near-nuclear density through localized one-center expansions.\cite{Lippert1999}  The smooth contribution is treated using the \GPW{} machinery, while the localized contributions are evaluated using one-center integrals and atom-centered grids.\par With \texttt{POTENTIAL ALL}, the AO density matrix itself represents the all-electron state.  \GauXC{} therefore evaluates the complete all-electron AO density directly and does not call \Skala{} separately for the smooth, hard, and soft terms.  For the native PBE reference calculations, \texttt{GAPW\_ACCURATE\_XCINT} retains the complete augmented one-center XC integration in energies and derivatives.  This setting improves the native reference but does not alter the AO density matrix supplied to \GauXC{}.

\subsection{\GAPW{} with pseudopotentials and effective-core potentials}

The \GAPW{} machinery can also be combined with a valence-only Hamiltonian defined by a \GTH{} pseudopotential or an ECP.  In that case the internal smooth and one-center representations do not change the electron content of the AO density matrix: core electrons removed by the effective Hamiltonian remain absent.  The molecular \GauXC{} interface therefore passes the corresponding AO valence density directly to \Skala{} and does not add a separate one-center correction to the model energy.  This construction permits \GAPW{} molecular basis sets and def2/ECP protocols for heavier elements while retaining a well-defined valence-density input to \Skala{}.\cite{Weigend2005,Andrae1990}
\endgroup

\section{{\Skala{} Through \GauXC{}}}

\begingroup

The external-library evaluation separates \CPtwoK{}'s internal auxiliary density representations from the XC model.  \CPtwoK{} supplies the molecular geometry, Gaussian basis-set information, spin-resolved AO density matrices, and the MPI communicator.  \GauXC{} constructs and partitions the atom-centered molecular quadrature, evaluates either a conventional functional or \Skala{}, and returns the XC energy, AO XC potential matrix, and available nuclear derivatives.

\Skala{} is a learned enhancement-factor functional rather than a fixed semilocal expression.\cite{Skala2025,SkalaGitHub,SkalaModelCard}  On the molecular quadrature its energy can be written schematically as
\begin{equation}
  E_{\mathrm{xc}}^{\theta}
  =
  -\frac{3}{4}
  \left(\frac{6}{\pi}\right)^{1/3}
  \sum_{i=1}^{G}
  w_i
  \left(
    \rho_{\alpha,i}^{4/3}
    +
    \rho_{\beta,i}^{4/3}
  \right)
  f_{\theta}[\mathbf x]_i .
  \label{eq:skala_energy}
\end{equation}
Here, \(G\) is the number of quadrature points, \(w_i\) are the associated weights, and \(\theta\) denotes the model parameters.  The meta-generalized-gradient-approximation (meta-GGA)-like primitive feature vector is
\begin{equation}
  \begin{split}
    \mathbf x_i = \big(&
      \rho_{\alpha,i},
      \rho_{\beta,i},
      |\nabla\rho_{\alpha,i}|,
      |\nabla\rho_{\beta,i}|,\\[-2pt]
      &\tau_{\alpha,i},
      \tau_{\beta,i},
      |\nabla\rho_{\alpha,i}+\nabla\rho_{\beta,i}|\big).
  \end{split}
  \label{eq:skala_features}
\end{equation}
The feature vector contains the two spin densities, their gradients, and their positive Kohn--Sham kinetic-energy densities \(\tau_\alpha\) and \(\tau_\beta\).  These ingredients are semilocal, but the learned enhancement factor is not: the atom-partitioned quadrature organizes efficient communication through assigned coarse points, so projected atom-centered messages couple different quadrature points before the final enhancement factor is formed.  This nonlocal coupling is why the complete density representation must be fixed before feature construction.  In the present implementation, \GauXC{} owns the integration grid, coarse-point assignment, descriptor evaluation, neural-network evaluation, and back-propagated derivatives.  It constructs the primitive fields \(\rho\), \(\nabla\rho\), and \(\tau\) from one AO density matrix on one molecular quadrature and builds the model features only afterwards.  No separate smooth and one-center \Skala{} evaluations are combined.

The differentiable outputs required by \CPtwoK{} are
\begin{equation}
  E_{\mathrm{xc}}^{\Skala},
  \qquad
  V_{\mu\nu,\sigma}^{\Skala}
  =
  \frac{\partial E_{\mathrm{xc}}^{\Skala}}
       {\partial P_{\mu\nu}^{\sigma}},
  \qquad
  \mathbf g_A^{\Skala}
  =
  \frac{\partial E_{\mathrm{xc}}^{\Skala}}
       {\partial \mathbf R_A}.
  \label{eq:skala_outputs}
\end{equation}
They are evaluated inside \GauXC{} and returned through its C/Fortran interface.  Figure~\ref{fig:interface-schematic} summarizes the resulting division of responsibility.

\begin{figure*}[t]
  \centering
  \begin{tikzpicture}[
    box/.style={
      draw=blue!70!black,
      line width=0.8pt,
      rounded corners=2pt,
      align=center,
      minimum height=2.6cm,
      text width=3.25cm,
      inner sep=6pt
    },
    arrow/.style={-{Latex[length=2.2mm]},line width=0.9pt,draw=blue!70!black},
    node distance=0.75cm
  ]
    \node[box] (density) {
      \textbf{\CPtwoK{} density}\\[3pt]
      \GPWGTH{}: valence\\
      \GAPWAE{}: all-electron\\
      \GAPWGTH{}, \GAPWECP{}: valence
    };
    \node[box,right=of density] (matrix) {
      \textbf{AO representation}\\[3pt]
      \(P_{\mu\nu}^{\alpha}\), \(P_{\mu\nu}^{\beta}\)\\
      Gaussian basis\\
      geometry and communicator
    };
    \node[box,right=of matrix] (gauxc) {
      \textbf{\GauXC{} / \Skala{}}\\[3pt]
      atom-centered quadrature\\
      \(\rho,\nabla\rho,\tau\)\\
      nonlocal model evaluation
    };
    \node[box,right=of gauxc] (outputs) {
      \textbf{Returned to \CPtwoK{}}\\[3pt]
      \(E_{\mathrm{xc}}\)\\
      \(V_{\mu\nu,\sigma}^{\mathrm{xc}}\)\\
      \(\partial E_{\mathrm{xc}}/\partial\mathbf R_A\)
    };
    \draw[arrow] (density) -- (matrix);
    \draw[arrow] (matrix) -- (gauxc);
    \draw[arrow] (gauxc) -- (outputs);
  \end{tikzpicture}
  \caption{{Molecular \CPtwoK{}--\GauXC{}--\Skala{} interface.  The electronic Hamiltonian determines whether the AO density matrix represents the valence density of a pseudopotential or effective-core Hamiltonian or an all-electron density.  \GauXC{} constructs the molecular quadrature and model features and returns the XC energy, AO potential matrix, and nuclear derivatives to \CPtwoK{}.}}
  \label{fig:interface-schematic}
\end{figure*}
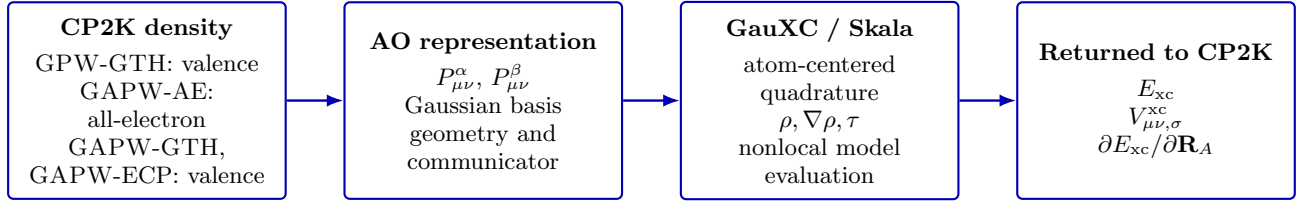
\endgroup

\section{{\CPtwoK{} Implementation}}

\begingroup

\CPtwoK{} stores AO matrices in distributed block-compressed sparse row (DBCSR) form,\cite{Schuett2016DBCSR} whereas the molecular \GauXC{} interface consumes and returns dense matrices.  Each MPI rank contributes its local density blocks, an all-reduction forms the replicated dense density matrix, and \GauXC{} evaluates \(E_{\mathrm{xc}}\) and \(V_{\mu\nu}^{\mathrm{xc}}\).  The returned matrix is symmetrized and inserted into a DBCSR matrix with the full upper-triangular block structure required by the quadrature result.  The remaining Hartree, external, pseudopotential, constraint, and basis-set contributions stay in their native \CPtwoK{} implementations.

All \GauXC{} grid and integrator objects are associated with the active MPI communicator, ensuring that density-matrix collection and quadrature evaluation remain local to communicator subgroups.  Compatible objects are reused across successive self-consistent-field (SCF) iterations.

For unrestricted calculations, \CPtwoK{} converts the spin-channel matrices to the scalar and collinear spin-density variables expected by \GauXC{} as follows:
\begin{equation}
  \mathbf P^{s}=\mathbf P^{\alpha}+\mathbf P^{\beta},
  \qquad
  \mathbf P^{z}=\mathbf P^{\alpha}-\mathbf P^{\beta},
  \label{eq:spin_transform}
\end{equation}
and transforms the returned derivatives back as \(\mathbf V^{\alpha}=\mathbf V^{s}+\mathbf V^{z}\) and \(\mathbf V^{\beta}=\mathbf V^{s}-\mathbf V^{z}\).
For \Skala{}, restricted and collinear-spin calculations share this density-variable convention.  The distinct one-spin normalization required by conventional restricted \GauXC{} kernels is applied only to a private working copy and does not alter either the spin-summed \CPtwoK{} density matrix or the \Skala{} input.

The nuclear derivative returned by \GauXC{} is an energy gradient, so the XC force contribution is
\begin{equation}
  \mathbf F_A^{\mathrm{xc}}
  =
  -\mathbf g_A^{\mathrm{xc}}
  =
  -\frac{\partial E_{\mathrm{xc}}}{\partial\mathbf R_A}.
  \label{eq:gauxc_force}
\end{equation}
For an isolated molecule, the same analytical gradient defines the force-based molecular virial around a fixed origin \(\mathbf R_0\):
\begin{equation}
  \Xi_{ij}^{\mathrm{mol}}
  =
  \sum_A
  g_{A,i}^{\mathrm{xc}}
  (R_{A,j}-R_{0,j})
  =
  -\sum_A
  F_{A,i}^{\mathrm{xc}}
  (R_{A,j}-R_{0,j}).
  \label{eq:molecular_virial}
\end{equation}
The scalar molecular virial reported below is \(W^{\mathrm{mol}}=\Tr[\boldsymbol\Xi^{\mathrm{mol}}]/3\).  It is a coordinate-scaling diagnostic for isolated molecules, not a periodic stress tensor.  The affine finite-difference deformations and validation conventions are documented in the Supplementary Information.
\endgroup

\section{Validation Protocol}

\begingroup

The computational protocol is organized around successive validation questions.  First, native \CPtwoK{} PBE and PBE evaluated through \GauXC{} are compared for the same density representation, geometry, basis, effective Hamiltonian, spin state, and numerical thresholds.  This isolates the AO density-matrix conversion, spin transformation, molecular quadrature, returned AO XC potential matrix, and force insertion from the model evaluation.  Second, \Skala{} energies are required to converge self-consistently, while analytical forces and the force-based molecular virial are compared with central finite differences of the total energy.

All calculations reported here use isolated-molecule boundary conditions.  The small-molecule set samples each AO density representation used by the interface, the water hexamer provides a focused many-body stress test, and dietGMTKN55 assesses the \Skala{} reference accuracy across chemically diverse reactions.  Exact basis sets, potentials, quadratures, real-space grids, SCF thresholds, finite-difference procedures, and benchmark-specific settings are documented in the corresponding sections of the Supplementary Information.

Table~\ref{tab:validation} gives one representative derivative check for each calculation class used in the manuscript.  The force errors compare one analytical component with the corresponding central finite-difference derivative and are reported in hartree/bohr.  The molecular-virial errors compare \(W^{\mathrm{mol}}\) with isotropic coordinate scaling and are reported in hartree.  Complete H$_2$, NH$_3$, H$_2$O, and HCl data, including all available PBE interface comparisons, are reported in Table S1.  The independent kinetic-energy-density validation with the Tao--Perdew--Staroverov--Scuseria (TPSS)\cite{Tao2003TPSS} and regularized--restored strongly constrained and appropriately normed (r$^2$SCAN)\cite{Furness2020R2SCAN} functionals at two grid resolutions is reported in Table S2.

\begin{table*}[t]
\caption{{Representative \Skala{} derivative checks for the molecular \GauXC{} interface.}}
\label{tab:validation}
\begin{ruledtabular}
\begin{tabular}{lllll}
System & \CPtwoK{} method & AO density supplied to \GauXC{} & \(|\Delta F_{\Skala{}-1.1}|\) & \(|\Delta W_{\Skala{}-1.1}|\)\\
\hline
H$_2$O & \GPWGTH{} & pseudopotential valence & \(9.17\times10^{-7}\) & \(1.09\times10^{-5}\)\\
H$_2$O & \GAPWAE{} & all-electron & \(1.99\times10^{-5}\) & \(7.25\times10^{-6}\)\\
H$_2$O & \GAPWGTH{} & pseudopotential valence & \(2.41\times10^{-6}\) & \(1.24\times10^{-5}\)\\
HCl & \GAPWECP{} & ECP valence & \(9.96\times10^{-6}\) & \(3.00\times10^{-6}\)\\
\end{tabular}
\end{ruledtabular}
\end{table*}
\endgroup

\section{{Complementary Water-Hexamer Stress Test}}

Before the broad benchmark, relative energies of eight neutral water-hexamer isomers provide a compact many-body test of cooperative polarization, exchange repulsion, charge redistribution, and nonlocal correlation.  For the primary \GAPWAE{} protocol, \Skala{}-1.1 reduces the mean unsigned error from 6.10 to 5.44~\(\mathrm{kcal\,mol^{-1}}\) relative to PBE, whereas the associated D3 dispersion correction with Becke--Johnson damping [D3(BJ)] does not improve the overall ordering.  Tables S3--S6 show comparable errors and ordering across density and core representations.  The nonmonotonic trend between the triple- and quadruple-zeta all-electron bases indicates partial basis-set--functional error cancellation.

\section{{Validation on dietGMTKN55}}

We evaluated all 100 reactions in dietGMTKN55\cite{Gould2018} and compared the reaction-energy errors obtained with \CPtwoK{}/\GauXC{} against a \Skala{} reference evaluation performed with \PySCF{}.\cite{Sun2020,Skala2025,SkalaGitHub}  The \CPtwoK{} calculations used \GAPWAE{} for elements up to bromine and \GAPWECP{} with def2 effective-core potentials for the heavier elements.  All \GAPWAE{}/\GAPWECP{} calculations in this primary set converged.  The resulting mean absolute deviation (MAD) is 1.255~\(\mathrm{kcal\,mol^{-1}}\), compared with 1.243~\(\mathrm{kcal\,mol^{-1}}\) for the \Skala{} reference evaluation, a difference of 0.012~\(\mathrm{kcal\,mol^{-1}}\).  The corresponding weighted total mean absolute deviations, type 2 (WTMAD-2),\cite{Gould2018} are 3.486 and 3.459~\(\mathrm{kcal\,mol^{-1}}\), respectively.

\begin{figure}[t]
  \centering
  \includegraphics[width=\columnwidth]
    {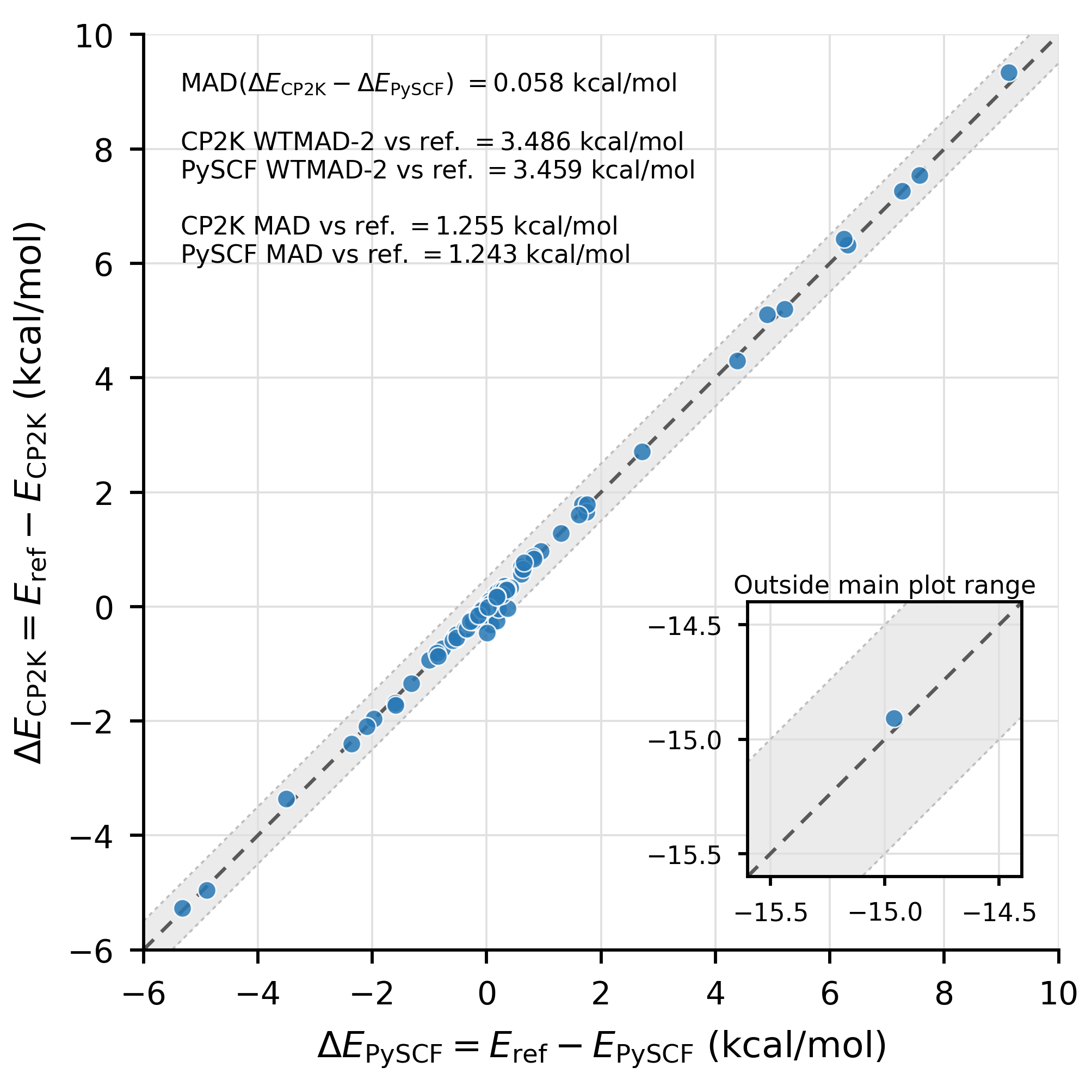}
  \caption{{Reaction-by-reaction correlation of the signed errors relative to the dietGMTKN55 reference energies obtained with \CPtwoK{}/\GauXC{} and the reference \Skala{} implementation.  The dashed line denotes equality, the shaded gray area a cross-code discrepancy of \(\pm0.5~\mathrm{kcal\,mol^{-1}}\), and the inset shows the only datapoint outside the main plot range.}}
  \label{fig:cp2k-pyscf-diet-accuracy}
\end{figure}

Figure~\ref{fig:cp2k-pyscf-diet-accuracy} shows that this agreement also holds reaction by reaction: the signed errors relative to the dietGMTKN55 reference energies have \(R^2=0.998\) between the two implementations.  The mean absolute direct cross-code residual is 0.058~\(\mathrm{kcal\,mol^{-1}}\), obtained by averaging \(\lvert\Delta E_{\mathrm{rxn}}^{\mathrm{CP2K/GauXC}}-\Delta E_{\mathrm{rxn}}^{\mathrm{PySCF/Skala}}\rvert\) over all 100 reactions.  This reaction-by-reaction measure is distinct from the 0.012~\(\mathrm{kcal\,mol^{-1}}\) difference between the two aggregate MADs.  All absolute cross-code residuals are below \(0.5~\mathrm{kcal\,mol^{-1}}\).  The remaining small residuals are consistent with differences in the numerical quadrature, grid partitioning, and SCF implementations.  For the primary \PySCF{}/\Skala{} comparison, the atomic-radius adjustment in the Becke partitioning was disabled; using \PySCF{}'s default Bragg-radii adjustment increases the cross-code deviations.  This sensitivity is documented in Fig.~S1, and the signed cross-code residuals are shown in Fig.~S2.

We additionally evaluated \Skala{} in the \GPW{} framework using the molecularly optimized triple-zeta valence with polarization TZVP-MOLOPT-SCAN-GTH basis and GTH-SCAN pseudopotentials.  This comparison changes both the basis set and the treatment of the core electrons and is therefore not a basis-set-only test against the \GAPWAE{}/\GAPWECP{} calculations above.  Of the 236 molecular calculations required for dietGMTKN55, 200 converged, yielding complete reaction energies for 75 of the 100 reactions.  The comparison in Table~\ref{tab:dietgmtkn55-molopt} excludes the anomalously basis-sensitive \(\mathrm{O_3\rightarrow3\,O}\) atomization reaction and therefore uses a common subset of 74 reactions.  WTMAD-2 is recomputed over this subset.  The full subset dependence and the SCF-convergence diagnostics are reported in Table S7 and Fig.~S3, respectively.

\begin{table}[t]
  \centering
  \caption{{MAD and WTMAD-2 values for the common 74-reaction subset after excluding the \(\mathrm{O_3\rightarrow3\,O}\) reaction.  All values are in \(\mathrm{kcal\,mol^{-1}}\).}}
  \label{tab:dietgmtkn55-molopt}
  \begin{ruledtabular}
  \begin{tabular}{lcc}
    Implementation & MAD & WTMAD-2 \\
    \hline
    \PySCF{}                & 0.875 & 3.869 \\
    \CPtwoK{} (\GAPW{})     & 0.865 & 3.878 \\
    \CPtwoK{} (\GPW{})      & 1.085 & 5.559 \\
  \end{tabular}
  \end{ruledtabular}
\end{table}

On this common subset, the \GPW{}/MOLOPT calculations have a MAD and WTMAD-2 of \(1.085\) and \(5.559~\mathrm{kcal\,mol^{-1}}\), respectively, compared with \(0.865\) and \(3.878~\mathrm{kcal\,mol^{-1}}\) for the corresponding \GAPW{}/def2 calculations (Table~\ref{tab:dietgmtkn55-molopt}).  The mean absolute difference between the two sets of predicted reaction energies is \(1.151~\mathrm{kcal\,mol^{-1}}\).  This pairwise quantity is distinct from either MAD relative to the dietGMTKN55 references.

For the reactions included in this comparison, the \GPW{} calculations are slightly less accurate than the def2-based \GAPW{} calculations, with a MAD larger by \(0.220~\mathrm{kcal\,mol^{-1}}\).  The more substantial limitation is robustness: approximately 15\% of the \GPW{} molecular SCF calculations did not converge.  Because \Skala{} was trained on all-electron densities for the elements in the first two rows of the periodic table, removing their core electrons with \GTH{} pseudopotentials may place some systems outside the density distribution represented in the training data.

\section{Computational Performance and Outlook}

The molecular \GauXC{} implementation defines a natural accelerator boundary because quadrature construction, descriptor evaluation, and the neural model reside in the external library.  A controlled single-rank benchmark on an NVIDIA GB10 system shows the expected size dependence: fixed initialization costs make GPU execution slower for the water monomer, whereas the water octamer reaches wall-time and Kohn--Sham-matrix speedups of 3.28 and 3.83, respectively.  Central processing unit (CPU) and GPU energies agree within \(1.5\times10^{-7}\)~hartree across the tested sizes.  Exact timings, host-memory measurements, numerical settings, and energy differences are reported in Table S8.

Extending the present AO-density interface to periodic systems would require periodic atom-centered quadrature, lattice-image handling, \(k\)-point-resolved density and XC matrices, and analytical cell derivatives in \GauXC{}.  A complementary native-grid \CPtwoK{} formulation, in which \CPtwoK{} owns the periodic density features, distribution, forces, and stress, lies outside the scope of this molecular study and will be reported separately.

\section{{Conclusions}}

{We have implemented a molecular \Skala{} interface in \CPtwoK{} through \GauXC{} using a spin-resolved AO density-matrix representation.  The same interface covers the valence density of \GPWGTH{}, the all-electron density of \GAPWAE{}, and the valence densities of \GAPWGTH{} and \GAPWECP{}.  \GauXC{} returns the XC energy, AO XC potential matrix, and nuclear derivatives, which \CPtwoK{} incorporates into its native Kohn--Sham, force, and molecular-virial machinery.}

{PBE-through-\GauXC{} comparisons and finite-difference checks establish the numerical consistency of the energy, potential, force, and molecular-virial paths.  The water-hexamer test provides a focused many-body diagnostic, while the final dietGMTKN55 benchmark reproduces the aggregate \Skala{} reference accuracy within 0.012~\(\mathrm{kcal\,mol^{-1}}\) and the reaction-by-reaction error profile with \(R^2=0.998\).  Current single-rank measurements additionally show that GPU acceleration becomes effective as the molecular workload grows.}

\section*{Supplementary Material}  See the supplementary material for complete small-molecule energy, force, and molecular-virial validation; kinetic-energy-density checks; water-hexamer benchmark details; dietGMTKN55 sensitivity analyses and convergence diagnostics; and the hardware timing diagnostic.  \begin{acknowledgments}
{The authors thank the AI for Science team at Microsoft Research for valuable discussions.  The authors also thank the \CPtwoK{} and \GauXC{} development communities.}  Part of the research was funded by the German Research Foundation (DFG) (project numbers 417590517/Collaborative Research Centre 1415 and 519869949).
\end{acknowledgments}

\section*{Data Availability}

Input files, selected raw \CPtwoK{} outputs, extraction scripts, checksums, source-version metadata, and processed benchmark tables are available in the companion repository \url{https://github.com/DCM-Uni-Paderborn/Molecular-Skala-in-CP2K}.  The repository contains the molecular validation data for the \GauXC{}/\Skala{} interface and the \GAPW{} force and molecular-virial calculations discussed in the Supplementary Information.  The water-hexamer structures and reference binding energies are taken from the updated Benchmark Energy and Geometry Database (BEGDB) water-cluster data set,\cite{Rezac2008BEGDB,Manna2017WaterClusters} and the repository contains the processed binding and relative-energy tables reported here.  The exact \CPtwoK{} release, \GauXC{}/\Skala{} stack, and \textsc{MOLOPT\_UZH} basis versions used for the reported calculations are documented with the inputs.

\begin{samepage}
\section*{Use of AI-assisted tools}
During the preparation of this work, the authors used OpenAI ChatGPT/Codex (GPT-5.6 Sol) to support language editing, mathematical exposition, the development of data-analysis and plotting scripts, consistency checks, and LaTeX preparation. The authors reviewed and edited the resulting content and take full responsibility for the content of the published article.
\par
\end{samepage}

\bibliography{references}

\clearpage
\onecolumngrid
\setcounter{section}{0}
\setcounter{table}{0}
\setcounter{equation}{0}
\setcounter{figure}{0}
\renewcommand{\thesection}{S\arabic{section}}
\renewcommand{\thetable}{S\arabic{table}}
\renewcommand{\theequation}{S\arabic{equation}}
\renewcommand{\thefigure}{S\arabic{figure}}
\renewcommand{\theHsection}{supplement.\arabic{section}}
\renewcommand{\theHtable}{supplement.\arabic{table}}
\renewcommand{\theHequation}{supplement.\arabic{equation}}
\renewcommand{\theHfigure}{supplement.\arabic{figure}}
\begin{center}
{\large Supplementary Information for: Molecular Implementation of the Machine-Learned \Skala{} Exchange--Correlation Functional in \CPtwoK{} through \GauXC{}}\\[1ex]
Franz P\"oschel, Johann Pototschnig, Frederick Stein, Andreas Kn\"upfer, Thijs Vogels, Stefano Battaglia, Sebastian Ehlert, J\"urg Hutter, and Thomas D. K\"uhne\\[0.5ex]
Franz P\"oschel and Johann Pototschnig contributed equally to this work.\\[0.5ex]
Author affiliations are given in the main manuscript.
\end{center}

\section{{Small-Molecule Validation}}

\begingroup

All small-molecule validation calculations use isolated molecular boundary conditions in \CPtwoK{}/\textsc{Quickstep}.  The method labels follow the main manuscript.  The Gaussian and plane-wave (\GPW{}) method with Goedecker--Teter--Hutter (\GTH{}) pseudopotentials is denoted \GPWGTH{}.  All-electron Gaussian augmented plane-wave (\GAPW{}) calculations are denoted \GAPWAE{}, while pseudopotential \GAPW{} calculations with either \GTH{} pseudopotentials or molecular effective-core potentials (ECPs) are denoted \GAPWGTH{} and \GAPWECP{}, respectively.  The calculations follow the \textsc{MOLOPT\_UZH} protocol\cite{Mirhosseini2026UZH} with \(E_{\mathrm{cut}}=600\) Ry and \(E_{\mathrm{rel}}=60\) Ry.  \GPWGTH{} uses molecularly optimized triple-zeta valence basis sets with two sets of polarization functions (TZV2P) and matching Perdew--Burke--Ernzerhof (PBE) \GTH{} pseudopotentials.  \GAPWAE{} uses all-electron quadruple-zeta valence basis sets with two sets of polarization functions (QZVPP) from the \textsc{MOLOPT\_UZH} family, \texttt{POTENTIAL ALL}, and \texttt{GAPW\_ACCURATE\_XCINT} for the native PBE exchange--correlation (XC) reference.  \GAPWGTH{} and \GAPWECP{} use the atomic-orbital (AO) valence density of their corresponding effective Hamiltonians.  HCl provides the \GAPWECP{} diagnostic.

Within each native-PBE/PBE-through-\GauXC{} pair, the geometry, basis, effective Hamiltonian, spin convention, self-consistent-field (SCF) threshold, and numerical grid are identical.  Unless stated otherwise, energy differences are reported in hartree, force errors in hartree/bohr, and molecular-virial errors in hartree.
\endgroup

\begingroup

Forces are validated against central finite differences (FD) of the total energy according to
\begin{equation}
  F_{A,i}^{\mathrm{FD}}
  =
  -
  \frac{E(R_{A,i}+h)-E(R_{A,i}-h)}{2h}.
  \label{eq:si_force_fd}
\end{equation}
The displacement is \(h=10^{-3}\)~\AA{} for every reported force check.  The tested component is the \(z\) component on atom 2 for H$_2$, NH$_3$, and HCl and the \(y\) component on atom 2 for H$_2$O.

For the molecular virial, coordinates are deformed affinely about a fixed center \(\mathbf R_0\) according to
\begin{equation}
  R_{A,k}^{(\pm;ij)}
  =
  R_{0,k}
  +
  (R_{A,k}-R_{0,k})
  \pm
  \epsilon\,\delta_{ki}(R_{A,j}-R_{0,j}).
  \label{eq:si_affine}
\end{equation}
Each tensor component is then checked using
\begin{equation}
  \Xi_{ij}^{\mathrm{FD}}
  =
  \frac{E^{(+;ij)}-E^{(-;ij)}}{2\epsilon}.
  \label{eq:si_virial_tensor_fd}
\end{equation}
The scalar values tabulated below use isotropic coordinate scaling with the dimensionless strain \(\epsilon=10^{-4}\) and \(W^{\mathrm{mol}}=\mathrm{Tr}[\boldsymbol\Xi^{\mathrm{mol}}]/3\).  These molecular coordinate-scaling diagnostics must not be interpreted as periodic cell stresses.
\endgroup

\begingroup

Table~\ref{tab:si_validation_full} collects the complete molecular diagnostics underlying Table I of the main manuscript.  The energy difference is \(\Delta E_{\mathrm{PBE}}=E_{\mathrm{PBE}}^{\GauXC{}}-E_{\mathrm{PBE}}^{\CPtwoK{}}\).  The force and molecular-virial columns report absolute analytical--FD differences.  The table gives the PBE interface comparison and the PBE and \Skala{} derivative errors for the H$_2$, NH$_3$, H$_2$O, and HCl calculations.  Individual \GAPWGTH{} and \GAPWECP{} total energies are not compared across effective Hamiltonians because their energy zero depends on the chosen pseudopotential or effective-core potential.  The transferable validation quantities are matched native/external-library energy differences and derivatives.

\begin{table}[htbp]
\centering
\caption{{Complete small-molecule energy and derivative validation.}}
\label{tab:si_validation_full}
\SITableFont
\setlength{\tabcolsep}{3pt}
\begin{tabular}{@{}llrrrrr@{}}
\hline
System & Method & \(\Delta E_{\mathrm{PBE}}\) & \(|\Delta F_{\mathrm{PBE}}|\) & \(|\Delta W_{\mathrm{PBE}}|\) & \(|\Delta F_{\Skala{}-1.1}|\) & \(|\Delta W_{\Skala{}-1.1}|\)\\
\hline
H$_2$ & \GPWGTH{} & \(1.27\times10^{-7}\) & \(7.60\times10^{-7}\) & \(3.41\times10^{-9}\) & \(2.60\times10^{-7}\) & \(3.01\times10^{-6}\)\\
NH$_3$ & \GPWGTH{} & \(1.08\times10^{-5}\) & \(7.60\times10^{-8}\) & \(1.97\times10^{-7}\) & \(9.86\times10^{-8}\) & \(5.07\times10^{-6}\)\\
H$_2$O & \GPWGTH{} & \(8.93\times10^{-5}\) & \(4.69\times10^{-7}\) & \(4.69\times10^{-7}\) & \(9.17\times10^{-7}\) & \(1.09\times10^{-5}\)\\
H$_2$ & \GAPWGTH{} & \(1.74\times10^{-7}\) & \(7.60\times10^{-7}\) & \(3.41\times10^{-9}\) & \(6.44\times10^{-7}\) & \(3.21\times10^{-6}\)\\
NH$_3$ & \GAPWGTH{} & \(5.76\times10^{-6}\) & \(1.23\times10^{-7}\) & \(2.03\times10^{-7}\) & \(5.27\times10^{-8}\) & \(6.45\times10^{-6}\)\\
H$_2$O & \GAPWGTH{} & \(1.13\times10^{-4}\) & \(4.69\times10^{-7}\) & \(1.43\times10^{-7}\) & \(2.41\times10^{-6}\) & \(1.24\times10^{-5}\)\\
H$_2$ & \GAPWAE{} & \(1.11\times10^{-6}\) & \(7.65\times10^{-7}\) & \(3.56\times10^{-9}\) & \(9.47\times10^{-7}\) & \(1.63\times10^{-6}\)\\
NH$_3$ & \GAPWAE{} & \(-1.47\times10^{-5}\) & \(7.52\times10^{-8}\) & \(1.73\times10^{-7}\) & \(8.23\times10^{-6}\) & \(3.40\times10^{-5}\)\\
H$_2$O & \GAPWAE{} & \(-2.94\times10^{-5}\) & \(4.70\times10^{-7}\) & \(3.78\times10^{-7}\) & \(1.99\times10^{-5}\) & \(7.25\times10^{-6}\)\\
HCl & \GAPWECP{} & \(2.26\times10^{-5}\) & \(5.38\times10^{-7}\) & \(8.13\times10^{-9}\) & \(9.96\times10^{-6}\) & \(3.00\times10^{-6}\)\\
\hline
\end{tabular}
\end{table}
\endgroup

\section{{Kinetic-Energy-Density Validation}}

\begingroup

The \Skala{} descriptor path depends explicitly on the positive Kohn--Sham (KS) kinetic-energy density \(\tau\).  We therefore validate this ingredient independently for closed-shell H$_2$O using the Tao--Perdew--Staroverov--Scuseria (TPSS)\cite{Tao2003TPSS} and regularized--restored strongly constrained and appropriately normed (r$^2$SCAN)\cite{Furness2020R2SCAN} meta-generalized-gradient approximations.  Table~\ref{tab:si_tau_mgga_convergence} compares the standard \(600/60\) Ry grid with a tighter \(1200/80\) Ry grid.  The energy difference is \(\Delta E=E^{\GauXC{}}-E^{\CPtwoK{}}\).  Force errors are reported in hartree/bohr and molecular-virial errors in hartree.  The subscripts \(\CPtwoK{}\) and \(\GauXC{}\) identify analytical--FD errors from the native and external-library derivative paths, respectively, while \(\lvert\Delta W_{\mathrm{xc}}\rvert_{\GauXC{}}\) isolates the XC contribution to the virial check.  For \GPWGTH{}, tightening the grid reduces the native-\CPtwoK{} force and molecular-virial FD errors by approximately two orders of magnitude, whereas the corresponding \GauXC{} derivative errors and XC-only virial residuals are already stable.  The \GAPWAE{} derivatives show substantially weaker grid sensitivity.  The comparison therefore identifies the dominant effect as the slower native-grid convergence of the \GPW{} representation for \(\tau\)-dependent functionals.  It does not imply monotonic convergence of every \GAPWAE{} energy difference.

\begin{table}[htbp]
\centering
\SITableFont
\setlength{\tabcolsep}{2pt}
\caption{{Closed-shell H$_2$O kinetic-energy-density convergence diagnostic on the standard and tight grids.}}
\label{tab:si_tau_mgga_convergence}
\begin{tabular}{@{}llrrrrrr@{}}
\hline
Method & Functional & \(\Delta E\) & \(|\Delta F|_{\CPtwoK{}}\) & \(|\Delta F|_{\GauXC{}}\) & \(|\Delta W|_{\CPtwoK{}}\) & \(|\Delta W|_{\GauXC{}}\) & \(|\Delta W_{\mathrm{xc}}|_{\GauXC{}}\)\\
\hline
\multicolumn{8}{l}{(a) Standard grid: \(E_{\mathrm{cut}}=600\) Ry and \(E_{\mathrm{rel}}=60\) Ry}\\
\hline
\GPWGTH{} & TPSS & \(-7.019{\times}10^{-4}\) & \(1.29{\times}10^{-5}\) & \(4.84{\times}10^{-7}\) & \(7.38{\times}10^{-5}\) & \(4.44{\times}10^{-7}\) & \(1.05{\times}10^{-8}\)\\
\GPWGTH{} & r$^2$SCAN & \(-1.805{\times}10^{-4}\) & \(1.23{\times}10^{-5}\) & \(4.88{\times}10^{-7}\) & \(5.46{\times}10^{-5}\) & \(1.78{\times}10^{-7}\) & \(9.92{\times}10^{-9}\)\\
\GAPWAE{} & TPSS & \(-5.222{\times}10^{-4}\) & \(5.14{\times}10^{-7}\) & \(5.19{\times}10^{-7}\) & \(3.35{\times}10^{-7}\) & \(3.21{\times}10^{-7}\) & \(1.15{\times}10^{-8}\)\\
\GAPWAE{} & r$^2$SCAN & \(+9.28{\times}10^{-5}\) & \(5.30{\times}10^{-7}\) & \(5.22{\times}10^{-7}\) & \(4.13{\times}10^{-8}\) & \(3.04{\times}10^{-8}\) & \(1.18{\times}10^{-8}\)\\
\hline
\multicolumn{8}{l}{(b) Tight grid: \(E_{\mathrm{cut}}=1200\) Ry and \(E_{\mathrm{rel}}=80\) Ry}\\
\hline
\GPWGTH{} & TPSS & \(-3.309{\times}10^{-4}\) & \(4.76{\times}10^{-7}\) & \(4.85{\times}10^{-7}\) & \(4.42{\times}10^{-7}\) & \(4.46{\times}10^{-7}\) & \(1.06{\times}10^{-8}\)\\
\GPWGTH{} & r$^2$SCAN & \(-8.230{\times}10^{-5}\) & \(5.02{\times}10^{-7}\) & \(4.89{\times}10^{-7}\) & \(9.52{\times}10^{-8}\) & \(1.80{\times}10^{-7}\) & \(9.82{\times}10^{-9}\)\\
\GAPWAE{} & TPSS & \(-4.064{\times}10^{-4}\) & \(5.14{\times}10^{-7}\) & \(5.18{\times}10^{-7}\) & \(3.40{\times}10^{-7}\) & \(3.22{\times}10^{-7}\) & \(1.18{\times}10^{-8}\)\\
\GAPWAE{} & r$^2$SCAN & \(+1.164{\times}10^{-4}\) & \(5.17{\times}10^{-7}\) & \(5.22{\times}10^{-7}\) & \(5.55{\times}10^{-8}\) & \(3.15{\times}10^{-8}\) & \(1.19{\times}10^{-8}\)\\
\hline
\end{tabular}
\end{table}
\endgroup

\section{Water-Hexamer Benchmark Details}

\begingroup

The water-hexamer calculations provide a complementary many-body stress test rather than a second broad benchmark.  The low-lying prism, cage, book, bag, cyclic-chair, and cyclic-boat isomers probe small relative-energy splittings controlled by cooperative polarization, many-body exchange repulsion, charge redistribution, and dispersion-like nonlocal correlation.\cite{Xantheas2002,Santra2008,Bates2009,Manna2017WaterClusters}  The structures and coupled-cluster singles and doubles with perturbative triples [CCSD(T)] reference binding energies extrapolated to the complete-basis-set (CBS) limit without counterpoise corrections are taken from the updated water-cluster subset of the Benchmark Energy and Geometry Database (BEGDB) and are used without zero-point corrections.\cite{Rezac2008BEGDB,Manna2017WaterClusters}

The primary protocol uses \GAPWAE{}, selected with \texttt{POTENTIAL ALL}, QZVPP-quality \textsc{MOLOPT\_UZH} basis sets,\cite{Mirhosseini2026UZH} and \texttt{GAPW\_ACCURATE\_XCINT}.  The latter option retains the hard all-electron and soft compensation one-center contributions in the native PBE XC integration and therefore provides the appropriate native \CPtwoK{} comparison for the all-electron AO density evaluated by \GauXC{}.  PBE-through-\GauXC{} and native PBE use the same structures, basis, spin convention, SCF thresholds, and numerical grids.

The PBE-D3(BJ) values are native \CPtwoK{} calculations with Grimme's D3 dispersion correction and Becke--Johnson damping [D3(BJ)].\cite{Grimme2010DFTD3,Grimme2011D3BJ}  Following the public \Skala{}-1.1 model metadata, the \Skala{}-1.1-D3(BJ) values add the B3LYP5-D3(BJ) binding contribution as a separate post-SCF correction because the present interface keeps \texttt{XC\_FUNCTIONAL/\allowbreak GAUXC} and \texttt{VDW\_POTENTIAL} separate.

For isomer \(i\), the binding energy, relative energy, and \Skala{} relative-energy error are defined as follows:
\begin{align}
  E_{b,i}
  &=
  E_i^{(\mathrm H_2\mathrm O)_6}
  -
  6E^{\mathrm H_2\mathrm O},
  \label{eq:si_hexamer_binding}\\
  \Delta E_i^{\mathrm{hex}}
  &=
  E_{b,i}
  -
  \min_j E_{b,j},
  \label{eq:si_hexamer_relative}\\
  \delta_i^{\mathrm{hex}}
  &=
  \Delta E_{i,\Skala}^{\mathrm{hex}}
  -
  \Delta E_{i,\mathrm{ref}}^{\mathrm{hex}}.
  \label{eq:si_water_error}
\end{align}
Referencing each method to its own lowest-energy isomer removes absolute-energy offsets and focuses the comparison on the ordering within a fixed Hamiltonian and density representation.  The mean unsigned error (MUE) is the average of \(\lvert\delta_i^{\mathrm{hex}}\rvert\) over the eight isomers.  All relative energies and MUEs in Tables~\ref{tab:si_hexamer_qzvpp}--\ref{tab:si_hexamer_sensitivity} are reported in \(\mathrm{kcal\,mol^{-1}}\).

\begin{table}[htbp]
\centering
\SITableFont
\setlength{\tabcolsep}{2pt}
\caption{Primary \GAPWAE{} water-hexamer relative binding energies without zero-point corrections.}
\label{tab:si_hexamer_qzvpp}
\begin{tabular}{lrrrrrr}
\hline
Isomer & Ref. & \GauXC{}-PBE & PBE & PBE-D3(BJ) & \Skala{}-1.1 & \Skala{}-1.1-D3(BJ)\\
\hline
Prism & 0.00 & 0.55 & 0.54 & 0.33 & 0.37 & 0.06\\
Cage & 0.21 & 0.00 & 0.00 & 0.00 & 0.00 & 0.00\\
Book 1 & 0.63 & 10.26 & 10.25 & 11.10 & 9.82 & 11.13\\
Book 2 & 1.00 & 5.17 & 5.16 & 5.93 & 4.86 & 6.06\\
Cyclic chair & 1.54 & 16.21 & 16.21 & 17.90 & 14.92 & 17.52\\
Bag & 1.55 & 1.90 & 1.90 & 2.48 & 1.70 & 2.66\\
Cyclic boat 1 & 2.57 & 9.20 & 9.20 & 10.83 & 7.82 & 10.34\\
Cyclic boat 2 & 2.63 & 15.28 & 15.28 & 16.90 & 13.71 & 16.22\\
\hline
MUE & -- & 6.11 & 6.10 & 6.97 & 5.44 & 6.79\\
\hline
\end{tabular}
\end{table}

The \GauXC{}-PBE and native PBE columns agree at the level expected from the residual difference between the atom-centered \GauXC{} quadrature and the native augmented XC integration.  \Skala{}-1.1 reduces the MUE from 6.10 to \(5.44~\mathrm{kcal\,mol^{-1}}\), but it still overstabilizes the compact cage and prism family relative to the book and cyclic structures.  Adding the B3LYP5-D3(BJ) contribution improves the prism--cage splitting but increases the MUE over all eight isomers to \(6.79~\mathrm{kcal\,mol^{-1}}\).  These results are therefore interpreted as a demanding many-body stress test rather than as evidence of uniform improvement for hydrogen-bonded networks.

Table~\ref{tab:si_gapw_tzvpp_hexamer} gives the \GAPWAE{} triple-zeta valence double-polarization (TZVPP) basis-set check.  It reproduces the same qualitative pattern as the QZVPP protocol: close \GauXC{}-PBE/native-PBE agreement, a moderate reduction of the MUE from \Skala{}-1.1, and no uniform improvement from the additive D3(BJ) correction.

\begin{table}[htbp]
\centering
\SITableFont
\setlength{\tabcolsep}{3pt}
\caption{Auxiliary \GAPWAE{} TZVPP water-hexamer relative binding energies for basis-set sensitivity.}
\label{tab:si_gapw_tzvpp_hexamer}
\begin{tabular}{lrrrrrr}
\hline
Isomer & Ref. & \GauXC{}-PBE & PBE & PBE-D3(BJ) & \Skala{}-1.1 & \Skala{}-1.1-D3(BJ)\\
\hline
Prism & 0.00 & 0.78 & 0.77 & 0.56 & 0.64 & 0.33\\
Cage & 0.21 & 0.00 & 0.00 & 0.00 & 0.00 & 0.00\\
Book 1 & 0.63 & 9.88 & 9.88 & 10.72 & 9.33 & 10.64\\
Book 2 & 1.00 & 4.72 & 4.72 & 5.48 & 4.29 & 5.50\\
Cyclic chair & 1.54 & 15.64 & 15.64 & 17.33 & 14.23 & 16.83\\
Bag & 1.55 & 1.41 & 1.40 & 1.99 & 1.09 & 2.05\\
Cyclic boat 1 & 2.57 & 8.76 & 8.76 & 10.39 & 7.35 & 9.87\\
Cyclic boat 2 & 2.63 & 14.86 & 14.86 & 16.48 & 13.25 & 15.76\\
\hline
MUE & -- & 5.83 & 5.83 & 6.66 & 5.17 & 6.41\\
\hline
\end{tabular}
\end{table}

Table~\ref{tab:si_gpw_hexamer} gives the complementary \GPWGTH{} calculation using PBE-optimized molecular basis sets and matching PBE \GTH{} pseudopotentials.  It documents the numerical behavior of the same \CPtwoK{}--\GauXC{} interface with an effective valence Hamiltonian.

\begin{table}[htbp]
\centering
\SITableFont
\setlength{\tabcolsep}{3pt}
\caption{Complementary \GPWGTH{} water-hexamer relative binding energies with PBE-optimized molecular basis sets and PBE \GTH{} pseudopotentials.}
\label{tab:si_gpw_hexamer}
\begin{tabular}{lrrrrrr}
\hline
Isomer & Ref. & \GauXC{}-PBE & PBE & PBE-D3(BJ) & \Skala{}-1.1 & \Skala{}-1.1-D3(BJ)\\
\hline
Prism & 0.00 & 1.18 & 1.11 & 0.90 & 0.92 & 0.61\\
Cage & 0.21 & 0.00 & 0.00 & 0.00 & 0.00 & 0.00\\
Book 1 & 0.63 & 10.30 & 10.29 & 11.14 & 10.10 & 11.41\\
Book 2 & 1.00 & 5.45 & 5.45 & 6.21 & 5.24 & 6.45\\
Cyclic chair & 1.54 & 15.55 & 15.42 & 17.11 & 14.60 & 17.20\\
Bag & 1.55 & 2.21 & 2.20 & 2.78 & 2.10 & 3.06\\
Cyclic boat 1 & 2.57 & 8.91 & 8.84 & 10.47 & 7.92 & 10.45\\
Cyclic boat 2 & 2.63 & 14.97 & 15.03 & 16.65 & 13.77 & 16.28\\
\hline
MUE & -- & 6.11 & 6.08 & 6.94 & 5.62 & 6.97\\
\hline
\end{tabular}
\end{table}

Table~\ref{tab:si_hexamer_sensitivity} summarizes the protocol sensitivity.  The QZVPP \GAPWAE{} row corresponds to the primary data in Table~\ref{tab:si_hexamer_qzvpp}.  The TZVPP \GAPWAE{} protocol probes basis-set sensitivity, and the \GPWGTH{} protocol probes the same workflow in a pseudopotential representation.  The maximum PBE mismatch is the largest absolute difference between \GauXC{}-PBE and native \CPtwoK{} PBE relative energies over the eight isomers.

\begin{table}[htbp]
\centering
\SITableFont
\setlength{\tabcolsep}{3pt}
\caption{Water-hexamer protocol sensitivity across the three density and basis representations.}
\label{tab:si_hexamer_sensitivity}
\begin{tabular}{lrrrrrr}
\hline
Protocol & \(\max|\Delta E_{\mathrm{PBE}}|\) & \GauXC{}-PBE & PBE & PBE-D3(BJ) & \Skala{}-1.1 & \Skala{}-1.1-D3(BJ)\\
\hline
\GAPWAE{} QZVPP & 0.009 & 6.11 & 6.10 & 6.97 & 5.44 & 6.79\\
\GAPWAE{} TZVPP & 0.008 & 5.83 & 5.83 & 6.66 & 5.17 & 6.41\\
\GPWGTH{} PBE & 0.125 & 6.11 & 6.08 & 6.94 & 5.62 & 6.97\\
\hline
\end{tabular}
\end{table}
\endgroup

\section{dietGMTKN55 Validation}

This section provides the numerical protocol and reaction-level diagnostics underlying Fig.~2 and Table~II of the main manuscript.  The dietGMTKN55 benchmark set is a representative subset of the General Main Group Thermochemistry, Kinetics, and Noncovalent Interactions (GMTKN55) database.\cite{Goerigk2017,Gould2018}  Single-point energies were evaluated at its fixed geometries.  The same \Skala{}-1.1 model was used in the reference calculations performed with \PySCF{} and, after conversion to the \GauXC{} model format, in \CPtwoK{}.  The calculations used spherical def2-TZVP basis functions~\cite{Weigend2005}.  The minimally augmented ma-def2-TZVP basis~\cite{Zheng2011} was selected for subsets that included anions.  All \Skala{} calculations included the D3(BJ) correction with B3LYP5 settings.\cite{Grimme2010DFTD3,Grimme2011D3BJ}  The mean absolute deviation (MAD) is the unweighted mean absolute reaction-energy error, whereas the weighted total mean absolute deviation, type 2 (WTMAD-2), combines subset-wise mean absolute deviations using the standard GMTKN55 size and energy-scale weights.\cite{Goerigk2017}

The \CPtwoK{} calculations used \CPtwoK{} 2026.2~\cite{CP2K2020}, with \GAPWAE{} for elements up to bromine and \GAPWECP{} for the heavier-element cases.  The \GauXC{} quadrature used the \texttt{FINE} grid, \texttt{ROBUST} pruning, and the Mura--Knowles radial quadrature.  The plane-wave cutoff and relative cutoff were 500 and 50~Ry, respectively, with four multigrid levels.  \texttt{EPS\_SCF} and \texttt{EPS\_DEFAULT} were \(10^{-5}\) and \(10^{-10}\), respectively, with a maximum of 100 SCF iterations.  The SCF equations were solved by diagonalization using direct density-matrix mixing with a mixing factor of 0.4 and two previous densities.  Each molecule was centered in a nonperiodic orthorhombic cell with at least \(15~\text{\AA}\) between the nearest atom and every cell face.  All calculations used the analytic Poisson solver.  Four reactions, comprising eight unique molecules containing Bi, Te, or I, used the corresponding def2 effective-core potentials.\cite{Metz2000,Peterson2003}

\begin{figure}[t]
  \centering
  \includegraphics[width=\textwidth]
    {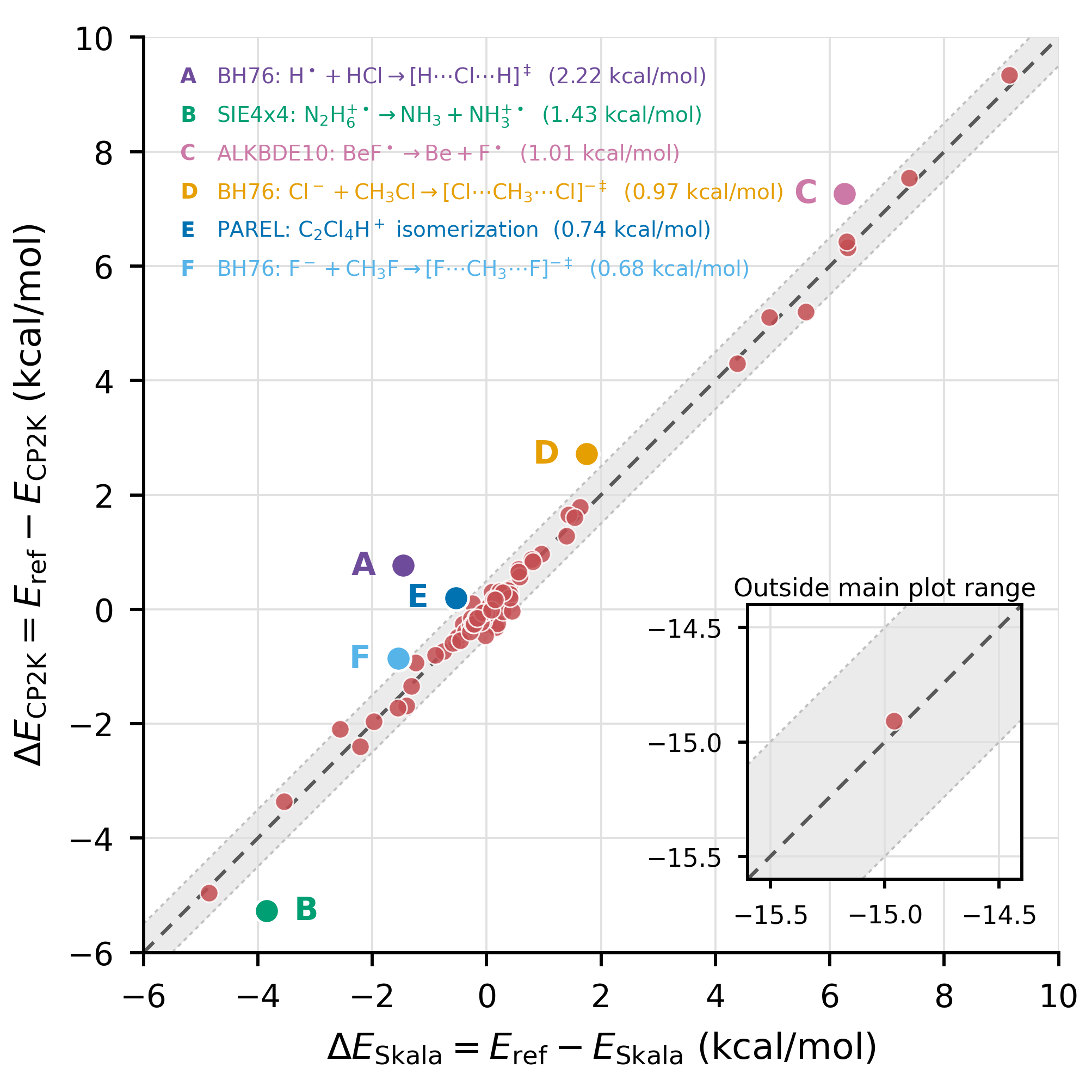}
  \caption{{Reaction-by-reaction correlation of the signed dietGMTKN55 errors obtained with \CPtwoK{}/\GauXC{} and the current open-source \Skala{} defaults.  The dashed line denotes equality, the shaded gray area a cross-code discrepancy of \(\pm0.5~\mathrm{kcal\,mol^{-1}}\), and the inset the only datapoint outside the main plot range.}}
  \label{fig:cp2k-skala-oss-diet-si}
\end{figure}

The primary reference calculations used \PySCF{} 2.10.0~\cite{Sun2020} with a level-3 Treutler--Ahlrichs atom-centered grid, original Becke partitioning \textbf{without} an atomic-radius adjustment, NWChem pruning, and a small-density cutoff of \(10^{-8}\).  The SCF energy and gradient thresholds were \(5\times10^{-6}~E_\mathrm{h}\) and \(10^{-3}\), respectively, with a maximum of 60 iterations.  Omitting the atomic-radius adjustment brings the partitioning convention into closer numerical correspondence with the \CPtwoK{}/\GauXC{} setup used for the primary comparison.  As a sensitivity test, we repeated the analysis with the \textit{default partitioning} of the current open-source \Skala{} distribution,\cite{SkalaGitHub} which shifts the Becke boundaries according to elemental Bragg radii.  Figure~\ref{fig:cp2k-skala-oss-diet-si} shows the resulting reaction-level comparison.  The changed partitioning increases the number of cross-code residuals larger than \(0.5~\mathrm{kcal\,mol^{-1}}\) from one to six, while leaving the aggregate statistics close to the primary result: its MAD and WTMAD-2 are \(1.228\) and \(3.512~\mathrm{kcal\,mol^{-1}}\), respectively.  Relative to \CPtwoK{}/\GauXC{}, these values differ by \(-0.027\) and \(+0.026~\mathrm{kcal\,mol^{-1}}\).

\begin{figure}[t]
  \centering
  \includegraphics[width=\textwidth]{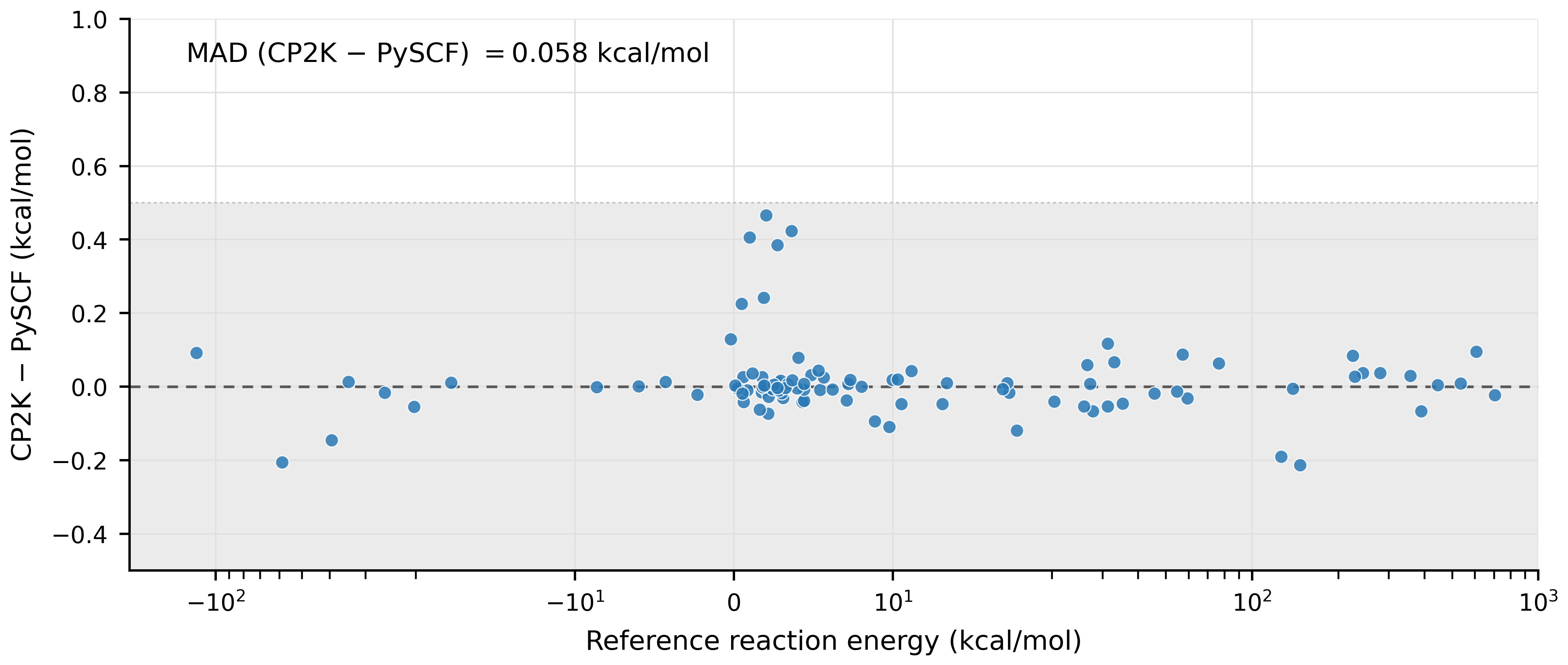}
  \caption{Reaction-by-reaction verification of the dietGMTKN55 energies.  Signed \CPtwoK{}/\GauXC{}--\PySCF{} reaction-energy differences are plotted against the reference reaction energies on a symmetric logarithmic horizontal axis; the interval from \(-10\) to \(+10~\mathrm{kcal\,mol^{-1}}\) is linear.}
  \label{fig:cp2k-pyscf-diet-si}
\end{figure}

Figure~\ref{fig:cp2k-pyscf-diet-si} resolves the primary cross-code difference directly against the reference reaction energy.  The mean absolute difference between the \CPtwoK{}/\GauXC{} and \PySCF{} reaction energies is \(0.058~\mathrm{kcal\,mol^{-1}}\).  All differences are smaller than \(0.5~\mathrm{kcal\,mol^{-1}}\), and the largest discrepancies are observed for reactions with the smallest reaction energies.

The additional \GPW{} calculations used the molecularly optimized triple-zeta valence with polarization TZVP-MOLOPT-SCAN-GTH basis and GTH-SCAN pseudopotentials, with a plane-wave cutoff of 1000~Ry and a relative cutoff of 60~Ry.  The \Skala{} model, D3(BJ) correction, \GauXC{} quadrature, molecular cells, Poisson treatment, and SCF settings were otherwise unchanged from the \GAPW{} calculations.  No additional diffuse functions were used for anionic reactions in this \GPW{} evaluation.  Of the 236 molecular calculations required for the 100 dietGMTKN55 reactions, 200 converged, yielding complete reaction energies for 75 reactions.

The largest sensitivity within the converged subset is the W4-11 \(\mathrm{O_3 \rightarrow 3\,O}\) atomization reaction.  Its reference energy is \(147.43~\mathrm{kcal\,mol^{-1}}\), whereas the \GAPW{} and \GPW{} predictions have errors of \(9.34\) and \(67.49~\mathrm{kcal\,mol^{-1}}\), respectively, and differ from one another by \(-58.15~\mathrm{kcal\,mol^{-1}}\).  Both calculations satisfy the nominal SCF convergence criterion.  Repeating the \(\mathrm{O_3}\) and atomic O calculations with the orbital-transformation SCF algorithm did not materially change the converged energies.  A similarly large discrepancy with r$^2$SCAN\cite{Furness2020R2SCAN} points to unusual basis sensitivity rather than a \Skala{}-specific effect.  The ozone reaction alone contributes \(0.90~\mathrm{kcal\,mol^{-1}}\) to the aggregate \GPW{} MAD.  After excluding it, the mean absolute \GPW{}--\GAPW{} reaction-energy difference is \(1.151~\mathrm{kcal\,mol^{-1}}\), and the mean signed difference is \(-0.081~\mathrm{kcal\,mol^{-1}}\).  Restricting the comparison further to the 65 reactions without anions gives reference MADs of \(0.976~\mathrm{kcal\,mol^{-1}}\) for \GPW{} and \(0.829~\mathrm{kcal\,mol^{-1}}\) for \GAPW{}, indicating that the remaining loss of accuracy is modest but somewhat larger for systems that used diffuse basis functions in the def2-TZVP evaluations.

The dependence of MAD and WTMAD-2 on the selected reaction subset is summarized in Table~\ref{tab:dietgmtkn55-molopt-si}.

\begin{table}[t]
  \centering
  \caption{MAD and WTMAD-2 values for the reaction subsets shared by the \PySCF{}, \GAPW{}, and \GPW{} calculations.  All values are in \(\mathrm{kcal\,mol^{-1}}\).}
  \label{tab:dietgmtkn55-molopt-si}
  \begin{tabular}{lccccccc}
    \hline
    & & \multicolumn{2}{c}{\PySCF{}} & \multicolumn{2}{c}{\CPtwoK{} (\GAPW{})} & \multicolumn{2}{c}{\CPtwoK{} (\GPW{})} \\
    Subset & \(N\) & MAD & WTMAD-2 & MAD & WTMAD-2 & MAD & WTMAD-2 \\
    \hline
    All common                              & 75 & 0.985 & 3.840 & 0.978 & 3.849 & 1.971 & 5.656 \\
    All common, excluding \(\mathrm{O_3}\)  & 74 & 0.875 & 3.869 & 0.865 & 3.878 & 1.085 & 5.559 \\
    No-diffuse functions                    & 66 & 0.952 & 4.104 & 0.958 & 4.106 & 1.984 & 5.900 \\
    No-diffuse, excluding \(\mathrm{O_3}\)  & 65 & 0.826 & 4.141 & 0.829 & 4.142 & 0.976 & 5.793 \\
    \hline
  \end{tabular}
\end{table}

\begin{figure}[t]
  \centering
  \includegraphics[width=\textwidth]{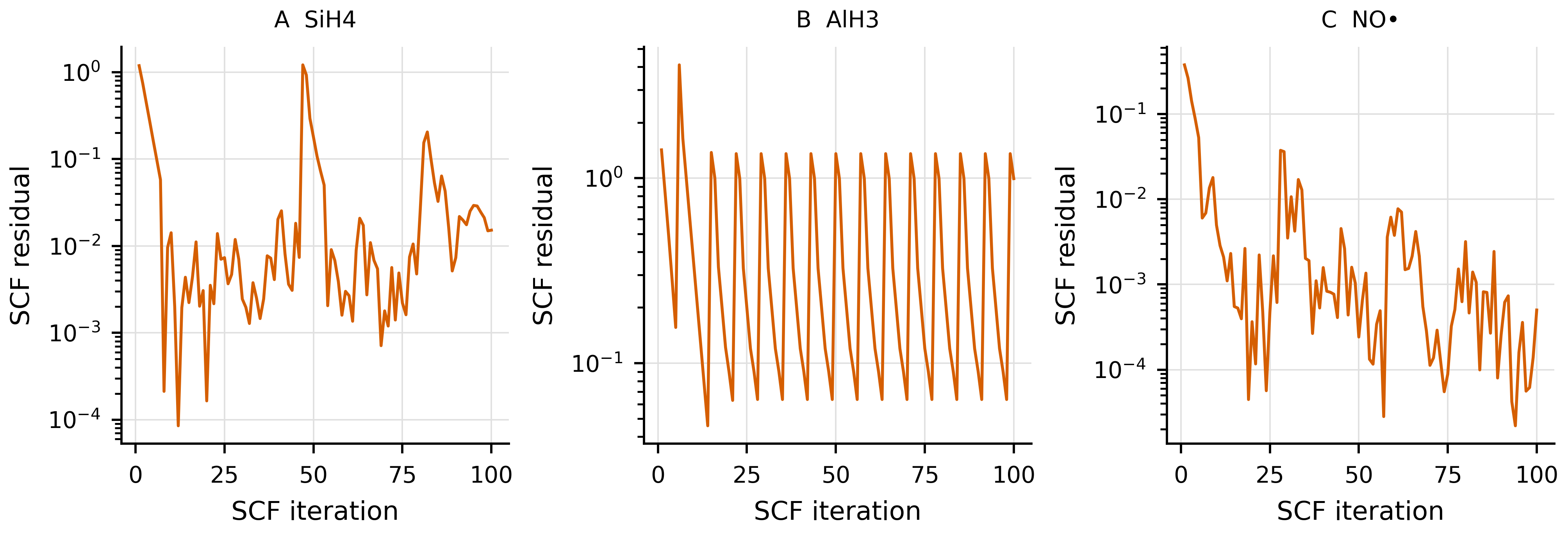}
  \caption{SCF residuals for representative unconverged \GPWGTH{} calculations.  The requested convergence threshold is \(10^{-5}\).}
  \label{fig:molopt-scf-failures}
\end{figure}

The 36 unconverged molecular calculations show unstable rather than uniformly slow convergence.  For representative \(\mathrm{SiH_4}\), \(\mathrm{AlH_3}\), and \(\mathrm{NO^\bullet}\) calculations, the lowest SCF residuals reached were \(8.51\times10^{-5}\), \(4.58\times10^{-2}\), and \(2.20\times10^{-5}\), respectively, while the final residuals were \(1.52\times10^{-2}\), \(9.93\times10^{-1}\), and \(5.01\times10^{-4}\).  Convergence required a residual at or below \(1.0\times10^{-5}\).  Thus, two examples approached the requested threshold before deteriorating, whereas the \(\mathrm{AlH_3}\) calculation remained far from convergence.  Because the model was trained on all-electron densities for first- and second-row elements, the valence-only \GPW{} density may lie outside the best-represented part of its training distribution for some systems.  This interpretation is consistent with, but not established uniquely by, the observed SCF behavior.  Figure~\ref{fig:molopt-scf-failures} shows the representative convergence curves, with A, B, and C denoting \(\mathrm{SiH_4}\), \(\mathrm{AlH_3}\), and \(\mathrm{NO^\bullet}\), respectively.

\clearpage
\section{{Hardware Timing Diagnostic}}

\begingroup

Table~\ref{tab:si_gpu_timing} reports a controlled molecular \GauXC{}/\Skala{} hardware-path diagnostic using \CPtwoK{} source revision \texttt{21ef8686db}, the \Skala{}-1.1 Rev1 host and CUDA-enabled model artifacts, and an NVIDIA GB10 system.  The isolated \GPWGTH{} water clusters were evaluated with one Message Passing Interface (MPI) rank and ten Open Multi-Processing (OpenMP) threads, one SCF KS-matrix build from an atomic guess, \(E_{\mathrm{cut}}=150\) Ry, \(E_{\mathrm{rel}}=30\) Ry, and the \GauXC{} \texttt{FINE}/\texttt{ROBUST} quadrature.  These deliberately lightweight numerical settings isolate the central processing unit (CPU) and graphics processing unit (GPU) execution paths and are not the production-accuracy protocol used for molecular validation.  One warm-up run and three measured runs were performed for each size and backend.  All times are reported in seconds, the speedup is \(S=t_{\mathrm{CPU}}/t_{\mathrm{GPU}}\), and the energy difference is \(\Delta E=E_{\mathrm{GPU}}-E_{\mathrm{CPU}}\).

\begin{table}[htbp]
\centering
\SITableFont
\setlength{\tabcolsep}{2pt}
\caption{Single-rank CPU/GPU \Skala{}-through-\GauXC{} timing diagnostic for isolated \((\mathrm{H}_2\mathrm{O})_n\) clusters on an NVIDIA GB10 system.  Times are medians over three measured runs after one warm-up run.}
\label{tab:si_gpu_timing}
\begin{tabular}{@{}rrrrrrrrr@{}}
\hline
\(n\) & Atoms & \(t_{\mathrm{wall}}^{\mathrm{CPU}}\) & \(t_{\mathrm{wall}}^{\mathrm{GPU}}\) & \(S_{\mathrm{wall}}\) & \(t_{\mathrm{KS}}^{\mathrm{CPU}}\) & \(t_{\mathrm{KS}}^{\mathrm{GPU}}\) & \(S_{\mathrm{KS}}\) & \(\Delta E\) (\(E_\mathrm{h}\))\\
\hline
1 & 3  & 3.37  & 3.72  & 0.91 & 1.89  & 1.99  & 0.95 & \(1.24{\times}10^{-8}\)\\
2 & 6  & 5.54  & 4.71  & 1.18 & 3.95  & 2.72  & 1.45 & \(2.94{\times}10^{-8}\)\\
4 & 12 & 13.42 & 5.73  & 2.34 & 11.62 & 3.58  & 3.24 & \(7.55{\times}10^{-8}\)\\
8 & 24 & 41.54 & 12.66 & 3.28 & 39.69 & 10.37 & 3.83 & \(1.48{\times}10^{-7}\)\\
\hline
\end{tabular}
\end{table}

The accelerator benefit grows with system size: the GPU is slower for the monomer because fixed initialization costs dominate, but reaches wall-time and Kohn--Sham-matrix speedups of 3.28 and 3.83, respectively, for the octamer.  The CPU/GPU energy difference remains below \(1.5\times10^{-7}\)~hartree for all four clusters.  The median peak host resident-set sizes for CPU/GPU execution are 1.94/2.84, 3.34/2.96, 5.77/3.06, and 9.95/3.09~GiB for \(n=1,2,4,8\), respectively.  The unified-memory GB10 platform does not expose a separate device-memory counter through \texttt{nvidia-smi}.  Consequently, only host resident-set size is reported.  These measurements quantify the single-rank hardware paths but do not constitute an MPI-scaling benchmark.
\endgroup

\clearpage

\end{document}